\documentclass[%
reprint,
superscriptaddress,
amsmath,amssymb,
aip,
longbibliography,
jap,
]{revtex4-1}
\usepackage{amsmath}
\usepackage{amssymb}
\usepackage{graphicx}
\usepackage{braket}
\usepackage{color}
\usepackage{xcolor}
\usepackage{siunitx}
\usepackage{upgreek}
\usepackage[utf8]{inputenc}
\usepackage{physics}
\usepackage{multirow}
\usepackage{array}
\usepackage{url}
\usepackage{dcolumn}
\usepackage{bm}
\usepackage{hyperref}
\usepackage{natbib}
\usepackage{setspace}
\usepackage{ulem}
\usepackage{placeins}

\begin{document}


\title{Electromagnetically induced transparency in inhomogeneously broadened divacancy defect ensembles in SiC}

\author{Olger V. Zwier*}
\affiliation{Zernike Institute for Advanced Materials, University of Groningen, NL-9474AG Groningen, the Netherlands}
\author{Tom Bosma*}
\affiliation{Zernike Institute for Advanced Materials, University of Groningen, NL-9474AG Groningen, the Netherlands}
\author{Carmem M. Gilardoni}
\affiliation{Zernike Institute for Advanced Materials, University of Groningen, NL-9474AG Groningen, the Netherlands}
\author{Xu Yang}
\affiliation{Zernike Institute for Advanced Materials, University of Groningen, NL-9474AG Groningen, the Netherlands}
\author{Alexander R. Onur}
\affiliation{Zernike Institute for Advanced Materials, University of Groningen, NL-9474AG Groningen, the Netherlands}
\author{Takeshi Ohshima}
\affiliation{National Institutes for Quantum Science and Technology, 1233 Watanuki, Takasaki, Gunma 370-1292, Japan}
\author{Nguyen T. Son}
\affiliation{Link\"{o}ping University, Department of Physics, Chemistry and Biology, S-581 83 Link\"{o}ping, Sweden}
\author{Caspar H. van der Wal}
\affiliation{Zernike Institute for Advanced Materials, University of Groningen, NL-9474AG Groningen, the Netherlands}

\date{Version of \today}


\begin{abstract}
	Electromagnetically induced transparency (EIT) is a phenomenon that can provide strong and robust interfacing between optical signals and quantum coherence of electronic spins. In its archetypical form, mainly explored with atomic media, it uses a (near-)homogeneous ensemble of three-level systems, in which two low-energy spin-1/2 levels are coupled to a common optically excited state. We investigate the implementation of EIT with c-axis divacancy color centers in silicon carbide. While this material has attractive properties for quantum device technologies with near-IR optics, implementing EIT is complicated by the inhomogeneous broadening of the optical transitions throughout the ensemble and the presence of multiple ground-state levels. These may lead to darkening of the ensemble upon resonant optical excitation. Here, we show that EIT can be established with high visibility also in this material platform upon careful design of the measurement geometry. Comparison of our experimental results with a model based on the Lindblad equations indicates that we can create coherences between different sets of two levels all-optically in these systems, with potential impact for RF-free quantum sensing applications. Our work provides understanding of EIT in multi-level systems with significant inhomogeneities, and our considerations are valid for a wide array of defects in semiconductors.
\end{abstract}

\maketitle


\section{Introduction}

The phenomenon of electromagnetically induced transparency (EIT) is well-studied for atomic media. This fundamentally quantum-mechanical phenomenon arises as an ensemble of three-level systems, each with two ground states laser coupled to a single optically excited state, evolves into a dark coherent superposition of ground-state sublevels (CPT state) \cite{fleischhauer2005,matsko2001}. As this happens, the absorption of the optical fields is suppressed, leading to a (partially) transparent medium. In this way, the phenomenon of EIT can be a resource that enables the preparation of coherent states all-optically without the need for additional microelectronic components. It allows for applications in areas such as sensing \cite{nagel1998,yudin2010,sedlacek2013,holloway2017,vafapour2017,Clevenson2015}, atomic clocks \cite{vanier2005}, low intensity non-linear optics \cite{harris1999,gorshkov2011,peyronel2012,firstenberg2013,chang2014}, topological photonics \cite{ozawa2019} and coherent photon storage for quantum memories \cite{duan2001,boyd2005,khurgin2005,kimble2008,lvovsky2009,novikova2012,gorniaczyk2014}.
When applied to ensembles, this technique further benefits from their collective behavior, leading to increased sensitivity and robust quantum-optical operation \cite{eisaman2005,acosta2013,lukin2003}.
Although many examples of near-perfect EIT exist for atomic ensembles \cite{gea1995,morigi2000,lukin2003,ma2017}, achieving this for solid-state systems is more challenging due to material inhomogeneities and interactions with the more complex environment. 
However, solid-state systems promise easier scalability and compatibility with existing technology platforms and may offer an advantage from the fact that quantum emitters in solids are at fixed locations in space. EIT can be established in solid-state systems by \textit{e.g.} using defect ensembles in crystal lattices as artificial atoms. Examples exist in rare-earth doped crystals \cite{turukhin2002,hedges2010}, donor-bound excitons in semiconductors \cite{sladkov2010} and the nitrogen-vacancy center in diamond \cite{acosta2013,doherty2013,Clevenson2015}. 

Here we show signatures of EIT in divacancy defect ensembles in silicon carbide. SiC is widely used in the semiconductor device industry and is compatible with silicon-based electronics, making SiC a promising platform for many electro-optical applications. Divacancies in SiC exhibit seconds-long spin-coherence times \cite{koehl2011, anderson2021} and allow mapping of electronic spin states onto nuclear spins \cite{seo2016,casas2017}, both of these at room temperature. 
Divacancy quantum emitters can be addressed in many ways. In recent years examples of optical control \cite{zwier2015}, charge-state control \cite{casas2017, Anderson2019}, and electrical and mechanical control \cite{falk2014, whiteley2019, klimov2014} have been demonstrated.
However, implementing EIT control for divacancies can still provide challenges.

Primarily, ensembles of divacancy defects have significant inhomogeneous broadening (orders of magnitude larger than the estimated lifetime-limited linewidths), even for the highest material quality currently envisioned \cite{waldermann2007,calusine2016,zhang2016,spindlberger2019,wolfowicz2020}. It is known that inhomogeneous broadening incurs in additional and stricter requirements on the intensity of the optical driving fields necessary for observing EIT \cite{gea1995}. Secondly, the divacancy contains a spin-triplet ground state. In these systems, repeated driving of optical transitions will usually lead to darkening of the ensemble due to optical pumping into dark spin sublevels. One way to avoid this is via the use of additional optical fields that counteract this optical pumping. Generally, such auxiliary fields add to the total dephasing of the system.

Here we provide an experimental check whether EIT of sufficient quality for quantum technologies can be obtained in this material system. In particular, we rely on the fact that the magnetic-field configuration can be engineered such that a single optical field drives transitions from two different ground-state spin sublevels\cite{zwier2015}. Additionally, we analyze our results in light of a model based on the master equations in Lindblad form, with inhomogeneous broadening taken into consideration. We use this analysis to extract relevant decay and dephasing parameters, and to show that, in certain conditions, the complicated multi-level system formed by ground and excited-state spin-sublevels can be reliably modeled by considering only a three-level system. We demonstrate that this approximation breaks down when multiple $\Lambda$ systems are driven simultaneously within the ensemble. We use this insight to engineer and experimentally demonstrate situations where double-EIT features appear, showing that we can prepare coherent states in these ensembles all-optically.

\section{Samples and methods}


We study EIT in an ensemble of c-axis divacancy spins in semi-insulating 4H-SiC. Divacancies were formed via electron irradiation and annealing after growth, to get an estimated divacancy concentration between 10$^{15}$ and 10$^{17}$ cm$^{-3}$. The sample was inserted in a liquid-helium flow cryostat with windows for optical access and equipped with a superconducting magnet system. This allowed us to apply a static magnetic field to tune the spin splittings of the $S=1$ ground state and the excited state. Further details on the sample preparation and optical measurements can be found in the Supplementary Information, Sec.~A and B. 

\begin{figure}[h!]
	\centering
	\includegraphics[width=8cm]{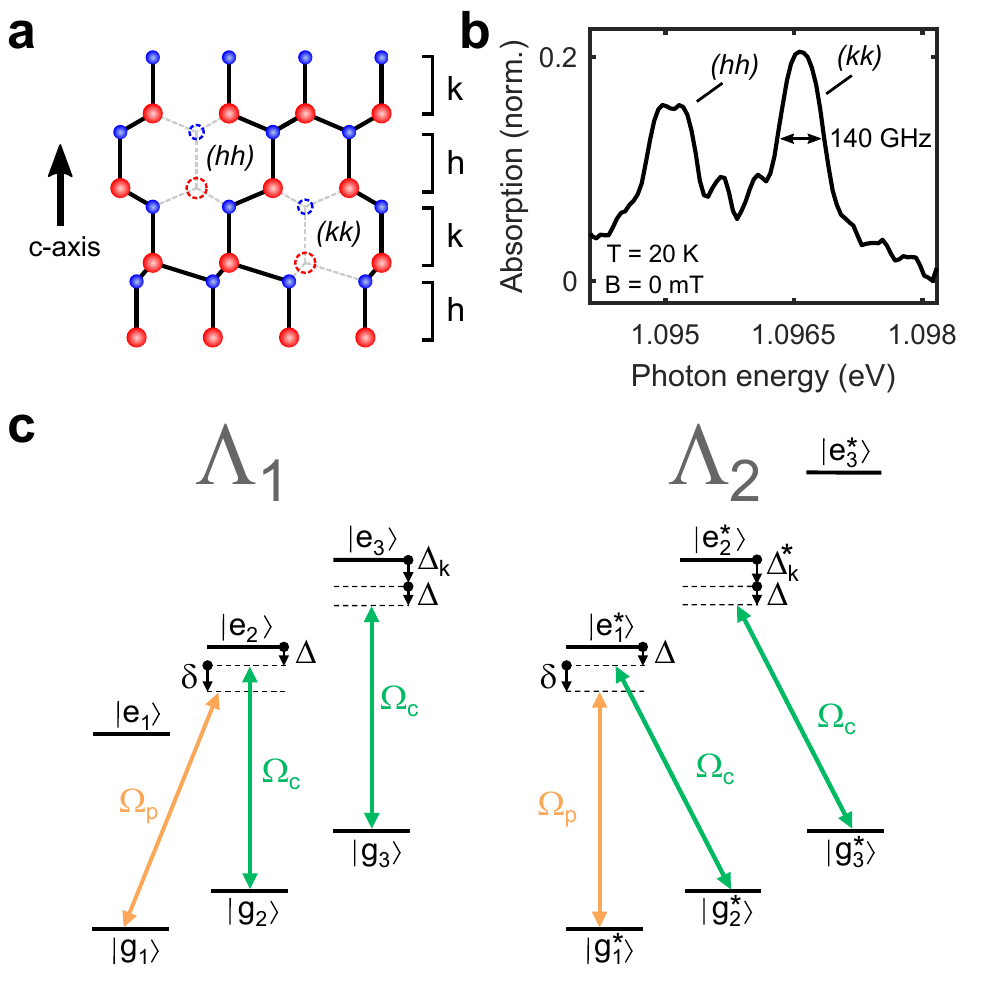}\\
	\caption{\textbf{The c-axis divacancies in 4H-SiC.} \textbf{a)} Schematic of the 4H-SiC lattice along the growth axis (c-axis), illustrating the occurrence of $(hh)$ and $(kk)$ c-axis divacancy defect centers. Ensembles of $(kk)$  divacancies are under study in this work. \textbf{b)} Result of scanning laser absorption spectroscopy revealing the zero-phonon lines of both c-axis divacancies. The $(kk)$ divacancy shows an inhomogeneous broadening of 140~GHz. \textbf{c)} Schematic of two $\Lambda$ schemes that occuring in our experiments. We choose to denote the spin sublevels by $\ket{g_i}$, $\ket{e_i}$ instead of the usual spin projections $\ket{m_s = 0, \pm1}$. Since we apply a magnetic field non-collinear with the symmetry axis of the defects (lattice-c-axis), the natural quantization axis along which the spin-projection is quantized differs for ground and excited states, and is along a non-trivial direction that does not conicide with either the magnetic field direction or the symmetry axis of the defect \cite{zwier2015,zwierThesis2016}. Our notation, however, only depends on the energy ordering of the states and is therefore more universal. The Lambda schemes can be driven such that optical spin pumping to a third ground-state level is avoided. Due to the large inhomogeneous broadening in our system both $\Lambda_1$ and $\Lambda_2$ are driven simultaneously in different subensembles. $\Omega_p$ and $\Omega_c$ denote the probe- and control-laser Rabi frequency, respectively. The control-laser detuning from the transition to the shared excited state is represented by $\Delta$, and the probe-laser detuning from two-laser resonance by $\delta$. Finally, $\Delta_k$ and $\Delta^*_k$ depict the detunings from the transitions to an additional excited state in $\Lambda_1$ and $\Lambda_2$, respectively.}
	\label{fig:EIT_FigPLE}
\end{figure}

The inhomogeneous broadening of the optical transition frequency can be revealed by scanning-laser absorption spectroscopy, where the sample absorption is measured as function of the frequency of an excitation laser. Figure~\ref{fig:EIT_FigPLE}b shows the resulting spectrum where the absorption is measured as a reduction in transmission of the excitation beam (calibrated on non-absorbing parts of the spectrum). The resolution of these measurements is limited by the linewidth of the laser, which is actively stabilized below a MHz. We present our results on a linear scale since the sample is only weakly absorbing (not optically thick). The laser is focused onto the sample to a spot of $70~\mu$m diameter. The Rayleigh length of the beam is comparable to the sample width (2 mm), such that the entire length of the sample contributes to the absorption signals. Two zero-phonon lines are visible: one at 1.0950~eV (1132 nm) related to divacancies at the \textit{(hh)} site \cite{koehl2011} (see Fig.~\ref{fig:EIT_FigPLE}a), and another at 1.0965~eV (1130 nm) related to divacancies at the \textit{(kk)} site. Both ZPLs associated with c-axis divacancies are sensitive only to excitation light polarized along the basal plane, that is, perpendicular to the symmetry axis of the defects \cite{falk2014}. Furthermore, both ZPLs are inhomogeneously broadened by 140~GHz. This value is a factor 5 broader than the ZPL width for as-grown divacancy ensembles in commercial wafers, and a factor 1000 broader than the lifetime-limited optical linewidth of these defects. This last ratio is close to 1 for atomic ensembles. \cite{gea1995,ma2017} In this work we focus on the c-axis divacancies at the \textit{(kk)} sites. At higher energies, several more ZPLs are present from \textit{(hk)} and \textit{(kh)} basal-plane divacancies (at 1.12~eV and 1.15~eV, respectively). These are not shown here \cite{falk2014}.


Figure~\ref{fig:EIT_FigPLE}c depicts our method for establishing EIT in the defect ensemble by coherent population trapping in the $\ket{g_1}$ and $\ket{g_2}$ spin levels (see the caption for an explanation of the notation $\ket{g_i}$, $\ket{e_i}$). As previously stated, signal loss due to optical spin pumping to the $\ket{g_3}$ level needs to be avoided. For an approach that does not rely on adding a third laser field this can be achieved by tuning to configurations where a single control laser simultaneously addresses two transitions close to resonance, \textit{i.e.} the $\ket{g_2}-\ket{e_2}$ and $\ket{g_3}-\ket{e_3}$ transitions in the left panel of Fig.~\ref{fig:EIT_FigPLE}c. This is enabled by static magnetic or electric fields, or strain \cite{falk2014, zwier2015}. In our case, we set the magnetic field angle $\varphi$ relative to the sample c-axis to $\ang{57}$ to achieve this. At this angle, two different optical transitions (from ground states $\ket{g_2}$ and $\ket{g_3}$) overlap for a broad range of magnetic field magnitudes \cite{zwierThesis2016} (see Supplementary Information, Sec.~C). In this configuration, the control laser depletes the populations of states $\ket{g_2}$ and $\ket{g_3}$, and prepares the population in the $\ket{g_1}$ spin state. A small misalignment of the magnetic field may cause an additional nonzero detuning $\Delta_k$ between the two different optical transitions driven by the control laser. Finally, a probe laser with variable frequency detuning relative to the control laser is used to excite from $\ket{g_1}$ to the shared excited state $\ket{e_2}$, completing the $\Lambda$ system and enabling the establishment of EIT upon two-photon resonance.

Both probe and control beams originate from the same laser. The detuning between them is obtained by passing the probe beam through an electro-optical modulator, such that they are intrinsically phase coherent and the detuning between them can be determined with sub-kHz accuracy. Additionally, we apply an off-resonance repump laser onto the sample to counteract slow ionization due to single-laser resonant excitation \cite{wolfowicz2017, casas2017, Anderson2019}. Details about the measurement setup can be found in the Supplementary Information, Sec.~B. Throughout this paper we use \textit{two-laser detuning} to indicate the frequency difference between the probe and control laser. Additionally, we use \textit{two-photon detuning $\delta$} to indicate the difference between the two-laser detuning and the two-photon resonance condition where EIT takes place.

The inhomogeneous broadening of the transition frequency causes the control-laser detuning $\Delta$ to vary for different defects within the ensemble. This leads to a situation where there is also a part of the ensemble that is closer to resonance with the $\Lambda_2$ system depicted in the right panel of Fig.~\ref{fig:EIT_FigPLE}c. This system has a different $\Delta_k^*$ value for the additional detuning between the two optical transitions being driven by the control laser. Both the $\Lambda_1$ and $\Lambda_2$ systems are always addressed simultaneously, albeit in different subensembles.


\section{EIT in two-laser spectroscopy}

\begin{figure}[h!]
	\centering
	\includegraphics[width=8cm]{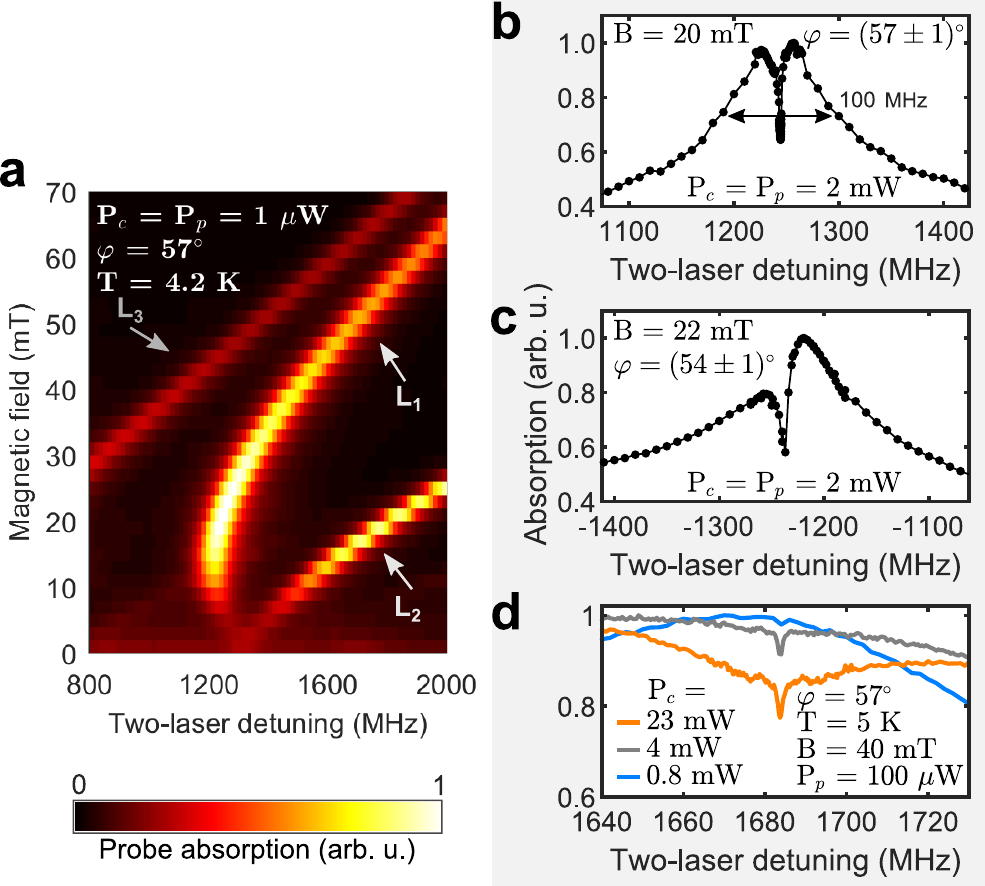}\\
	\caption{\textbf{EIT in two-laser absorption spectroscopy.} \textbf{a)} Two-laser magneto-spectroscopy scan at low laser powers revealing three clear absorption lines for the probe beam. Lines L\textsubscript{1} and L\textsubscript{2} correspond to $\Lambda$ driving schemes, whereas L\textsubscript{3} corresponds to a non-$\Lambda$ type of driving scheme. \textbf{b-d)} Absorption versus two-laser detuning for several scenarios: \textbf{b)} For a magnetic field angle $\varphi = 57^o$ that yields optimal $\Lambda$ system with $\Delta_k\approx0$ (L\textsubscript{1} in panel \textbf{a}), resulting in a symmetric EIT dip. \textbf{c)} For a slightly misaligned magnetic field $\varphi = 54^o$ where $\Delta_k \neq 0$, resulting in an asymmetric EIT dip. The dot-line traces in \textbf{b)}, \textbf{c)} represent experimental data. \textbf{d)} Control-beam power dependence of EIT in probe-beam absorption (for the $\Lambda$ system corresponding the L\textsubscript{1} in panel \textbf{a}). At larger control-beam power the EIT becomes more pronounced.}
	\label{fig:EIT_FigEIT}
\end{figure}

In order to identify the lambda systems we perform a two-laser magnetospectroscopy scan where the absorption is measured versus both two-laser detuning and magnetic-field strength for fields at angle $\varphi =\ang{57}$ with the c-axis. Note that throughout this work we always consider the absorption of the probe laser unless specified otherwise. The results are shown in Fig.~\ref{fig:EIT_FigEIT}a. In this range, three clear two-laser absorption features (TLAF) are visible with linewidths on the order of the homogeneous linewidth of the underlying transitions (and about a factor 1000 narrower than the inhomogeneous broadening of the ZPL). Upon relating the detuning and magnetic field values of the TLAF peaks to the energy eigenvalues of the theoretical effective-spin Hamiltonian (see the Supplementary Information, Sec.~C), we identify the features labeled L\textsubscript{1} and L\textsubscript{2} as resulting from two $\Lambda$ driving schemes, where two ground-state levels are coupled to one excited state. The L$_3$ feature results from a driving scheme where the lasers couple two ground-state spin sublevels to distinct excited-state levels, so EIT does not occur within this two-laser absorption feature (we checked for this experimentally).

Figure~\ref{fig:EIT_FigEIT}b shows a two-laser scan around the L\textsubscript{1} feature at higher resolution. A sharp EIT dip becomes visible in the center of the absorption peak. 
Figure~\ref{fig:EIT_FigEIT}c shows a similar result, but with the EIT feature appearing asymmetrically within the TLAF line. In the coming analysis, we will show that this occurs when the magnetic field direction deviates from the direction that gives optimal coincidence for the two different transitions being simultaneously driven by the control laser.

There is a well-known limit for the minimum control-laser Rabi frequency for complete EIT, \textit{i.e.} $\Omega_c^2 > \Gamma_e\gamma_g$ with $\Gamma_e$ the excited-state lifetime and $\gamma_g$ ground-state dephasing rate \cite{fleischhauer2005}. Here, we disregard the excited state dephasing time ($\gamma_e$). In EIT measurements, we cannot distinguish between excited-state optical decay processes and excited-state dephasing processes ($\Gamma_e$ and $\gamma_e$). Nonetheless, we will show in what follows that the excited-state dephasing processes are much slower than the optical decay processes at low temperatures, justifying our choice to disregard $\gamma_e$. Upon considering inhomogeneous broadening beyond $\Gamma_e$ the above-mentioned limit becomes \cite{gea1995}
\begin{equation}\label{eq:EITlimits}
	\Omega_c^2 > \Delta_I\gamma^*_g
\end{equation}
with $\Delta_I$ the FWHM inhomogeneous broadening. This limit was confirmed numerically via simulations \cite{zwierThesis2016}.
To test the power dependence of EIT in our inhomogeneously broadened system we scan around L\textsubscript{1} with several control-laser powers while keeping the probe-laser power fixed at 100~$\mu$W, much smaller than the control-laser power. The results are shown in Fig.~\ref{fig:EIT_FigEIT}d. The EIT dip becomes deeper at larger control-beam power, although the limit for complete EIT was not reached. Additionally, some power broadening of the TLAF and accompanying EIT dip is visible. 


\section{Model for inhomogeneous broadening}

The five-level $\Lambda_1$ system of Fig.~\ref{fig:EIT_FigPLE}c can often be accurately depicted by a three-level system with an additional ground-state relaxation term to account for decay from $\ket{g_1}$ or $\ket{g_2}$ to $\ket{g_3}$ and an additional excited-state decay term to account for the decay from the excited states towards the $\ket{g_3}$ level. We study the consequences of the large inhomogeneous broadening of the transition frequencies in our system on the EIT lineshape by considering such a three-level model. Note that we will show later that this conversion does not hold when the magnetic field is not perfectly aligned ($\Delta_k\neq0$) or when more than the three laser couplings in Fig.~\ref{fig:EIT_FigPLE}c occur.

For the $\Lambda$ system made up by $\ket{g_1}$, $\ket{g_2}$ and $\ket{e_2}$ (case $\Lambda_1$, Fig.~\ref{fig:EIT_FigPLE}c) we set up a Master equation in Lindblad form. We solve for the steady-state density matrix considering the parameters listed in table \ref{tab:EITmodelParams}. We assume that the decay rate $\Gamma_e$ from the excited state towards either ground state is equal.

\begin{table}
	\caption{\textbf{Parameter choices for three-level $\Lambda$ scheme model.}}
	\label{tab:EITmodelParams}
	\begin{center}
		\begin{tabular}{c | c} 
			parameter & value (Hz)\\\hline
			$\Gamma_e$ & $10^7$\\
			$\gamma_e$ & $\ll \Gamma_e$ \\
			$\Gamma_g$ & $10^4$\\
			$\gamma^*_g$ & $10^5$\\
			$\Omega_c$ & $3\times 10^6$\\
			$\Omega_p$ & $10^4$\\	
			$\Delta$ & variable\\
			$\delta$ & variable\\	
		\end{tabular}
	\end{center}
\end{table}

\begin{figure}[h!]
	\centering
	\includegraphics[width=8cm]{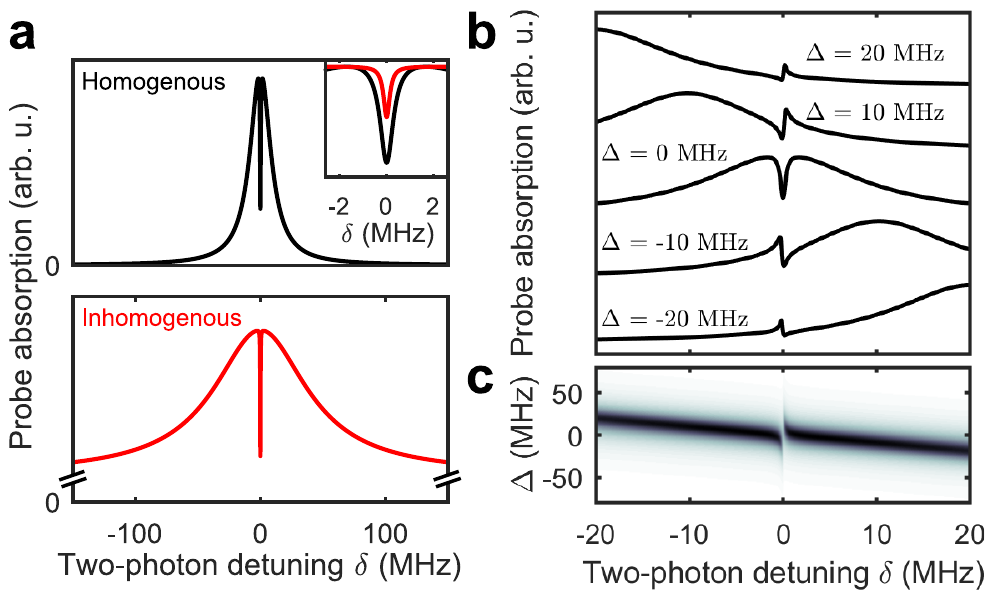}\\
	\caption{\textbf{Simulation results for the effect of inhomogeneous broadening on EIT lineshapes.} In an ensemble where the optical transition frequency is inhomogeneously broadened (Gaussian distribution) the control-laser detuning $\Delta$ is distributed accordingly. The total response of the probe absorption is then integrated over all defects with their respective detuning $\Delta$. \textbf{a)} Simulated EIT traces for (black) a homogeneous ensemble on resonance ($\Delta = 0$), and for (red) an inhomogeneously broadened ensemble (FWHM~=~100~GHz). 
The TLAF feature of the red trace appears on a broad background for the probe absorption (further discussed in the main text and Supplementary Information part~F).
The inset shows the EIT dip at higher resolution. Absorption traces are calculated for the case where the optical depth is below 1. \textbf{b)} Traces for homogeneous ensembles at various control-laser detunings $\Delta$ (vertical offset for clarity). At larger $\Delta$ the peak splits up in a linear probe-laser absorption peak at $\delta=-\Delta$, and a small absorption feature with both EIT and enhanced Raman characteristics close to two-photon resonance. A weighted summation of these traces for all $\Delta$ values yields the red curve in panel \textbf{a}. \textbf{c)} Homogeneous probe-laser absorption versus both control-laser detuning $\Delta$ and detuning from two-photon resonance $\delta$ (darker means increased absorption). The traces in panel \textbf{b} are horizontal cross-sections of this panel.}
	\label{fig:EIT_FigInhom}
\end{figure}

A comparison between theoretically obtained absorption spectra with EIT features for a homogeneous ensemble and an inhomogeneous one, where the optical transition is normally distributed with a FWHM of 100~GHz, is shown in Fig.~\ref{fig:EIT_FigInhom}a. The overall absorption peak for the inhomogeneous ensemble broadens and appears on top of a broad background for the probe absorption level.
This offset absorption is related to absorption from highly detuned subensembles, and only occurs outside the spectral range of the EIT dip (that is, it does not influence the depth of the EIT feature). Thus, even ensembles where this offset absorption is present will show full EIT with zero probe absorption at two-laser detuning $\delta = 0$ if the dephasing time is long enough. These aspects are further discussed in the Supplementary Information, part~F. Additionally, for the inhomogeneous ensemble, the EIT dip becomes shallower and narrower. This is clear from the EIT limits in Eq.~\ref{eq:EITlimits}, but we will discuss its origin in more detail in the next paragraph.

The fading of the EIT dip can be clarified by looking at the separate homogeneous subensembles that make up the inhomogeneous ensemble. Figure~\ref{fig:EIT_FigInhom}b shows probe-absorption traces for several subensembles with different values for the optical transition detuning $\Delta$. The detuning $\Delta$ is defined as zero when the control laser is resonant with the $\ket{g_2}-\ket{e_2}$ transition (see Fig.~\ref{fig:EIT_FigPLE}c). The inhomogeneous trace of Fig.~\ref{fig:EIT_FigInhom}a is the summation of these separate traces for all $\Delta$ values, weighted according to the inhomogeneous distribution of $\Delta$ values with FWHM of $100$~GHz.
For subensembles where $\Delta$ is nonzero but within the homogeneous linewidth $\Gamma_e$ the absorption peak with EIT dip becomes asymmetric, with an off-center dip. Nonetheless, for the full inhomogeneous ensemble of three-level systems the overall absorption peak will remain symmetric as there are equal amounts of blue- and red-detuned subensembles, averaging out the asymmetry. For larger optical-transition detuning $\Delta$ (larger than the homogeneous linewidth $\Gamma_e$) the lineshape splits up. Around $\delta = -\Delta$ a linear absorption peak (TLAF) appears as the probe laser becomes resonant with the $\ket{g_1}-\ket{e_2}$ transition. This gives rise to an offset in the TLAF of the inhomogeneous ensemble, when compared to the homogeneous one, in Fig.~\ref{fig:EIT_FigInhom}a. Simultaneously, close to two-photon resonance ($\delta = 0$) an asymmetric peak with both Raman and EIT characteristics appears \cite{fleischhauer2005}. Since this peak is typically narrower than the homogeneous EIT dip, the total inhomogeneous EIT dip is narrower than the homogeneous EIT dip (see Fig.~\ref{fig:EIT_FigInhom}a, inset). The tail of the Raman-EIT peak at $\delta = 0$ thus causes the shallower EIT dip for the inhomogeneous ensemble. In Fig.~\ref{fig:EIT_FigInhom}c the $\Delta$ dependence of the probe-absorption traces is shown at higher resolution (darker means higher absorption), revealing the evolution of the homogeneous absorption line along $\Delta$ and $\delta$.


\section{Model for asymmetric EIT}

As shown above, the appearance of off-center EIT is quite common and well understood for systems with small inhomogeneous broadening  \cite{gea1995}, where it arises from driving of detuned Raman transitions. In systems with large broadening, however, this asymmetry is usually averaged out when considering the response of the full ensemble. Nonetheless, we do observe off-center EIT dips in broad inhomogeneous ensembles, as in Fig.~\ref{fig:EIT_FigEIT}c. Below, we show that the appearance of these asymmetric EIT dips is related to imperfect overlap of the two optical transitions driven by the control laser ($\ket{g_2}-\ket{e_2}$ and $\ket{g_3}-\ket{e_3}$ transitions in the $\Lambda_1$ system in Fig.~\ref{fig:EIT_FigPLE}c).

\begin{figure}[h!]
	\centering
	\includegraphics[width=5cm]{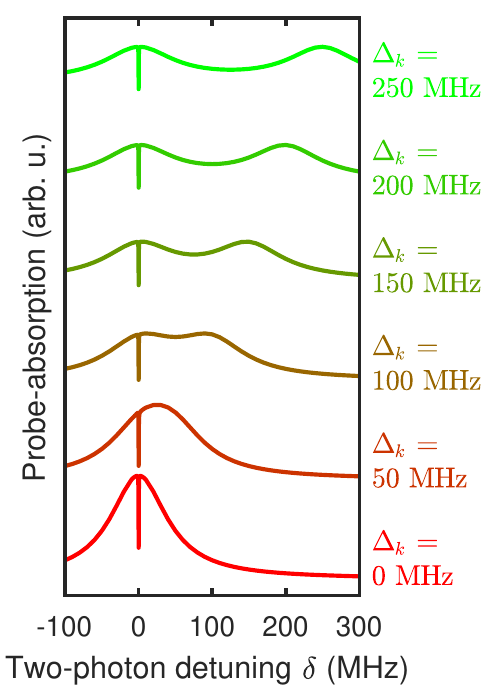}\\
	\caption{\textbf{Asymmetric EIT in five-level system.} Simulated probe-absorption traces for an inhomogeneously broadened ensemble (FWHM~=~100~GHz) at various $\Delta_k$ values (vertical offset for clarity). Imperfect matching of the $\ket{g_2}-\ket{e_2}$ and $\ket{g_3}-\ket{e_3}$ transition frequency defines the $\Delta_k$ value (see the $\Lambda_1$ scheme in Fig.~\ref{fig:EIT_FigPLE}c). For nonzero $\Delta_k$, there are two subensembles in the range of control-laser detuning values $\Delta$ where high probe-laser absorption occurs: One at $\Delta = 0$, yielding the absorption peak with EIT feature around $\delta = -\Delta = 0$, and another at $\Delta = -\Delta_k$, showing a linear absorption peak around $\delta = -\Delta = \Delta_k$. For low and nonzero $\Delta_k$ the total probe absorption thus becomes asymmetric.}
	\label{fig:EIT_FigDkAsymmetry}
\end{figure}

As a result of the imperfect overlap between the two optical transitions driven by the control laser, the $\Delta_k$ detuning becomes nonzero. In this case, two subensembles contribute to the enhanced absorption of a probe laser driving transitions between the ground state $\ket{g_1}$ and the optically excited states $\ket{e_2}$ and $\ket{e_3}$. These are the subensembles centered at $\Delta = 0$ and $\Delta = -\Delta_k$. For small nonzero $\Delta_k$ this causes an asymmetry in the TLAF lineshape as the contributions from each of the subensembles becomes unequal. This can be seen in Fig.~\ref{fig:EIT_FigDkAsymmetry}. Here, we show the inhomogeneous probe-laser absorption traces, as obtained from solving the Master equation for a five-level system, for several values of $\Delta_k$. At larger $\Delta_k$ the two peaks separate, resulting in an absorption peak around $\delta = 0$ with an on-center EIT dip, and a linear absorption peak around $\delta = \Delta$. The latter shows no EIT feature, since this absorption peak is no longer related to a $\Lambda$ driving scheme. 
For cases where $\Delta_k$ is comparable to the homogeneous linewidth $\Gamma_e$ the TLAF lineshape becomes asymmetric with an off-center EIT dip.


\section{Double EIT in five-level system}

Insofar, we have not discussed the situation that arises when probe and control fields drive, within the same subensemble, two different sets of $\Lambda$ schemes. As we show in what follows, this situation gives rise to two independent EIT dips within the same TLAF width. Each of these EIT dips corresponds to coherently trapping the spin population in different superpositions of ground-state levels, demonstrating that we can create ground-state coherences between different pairs of ground-state spin sublevels in these ensembles.

\begin{figure}[h!]
	\centering
	\includegraphics[width=8cm]{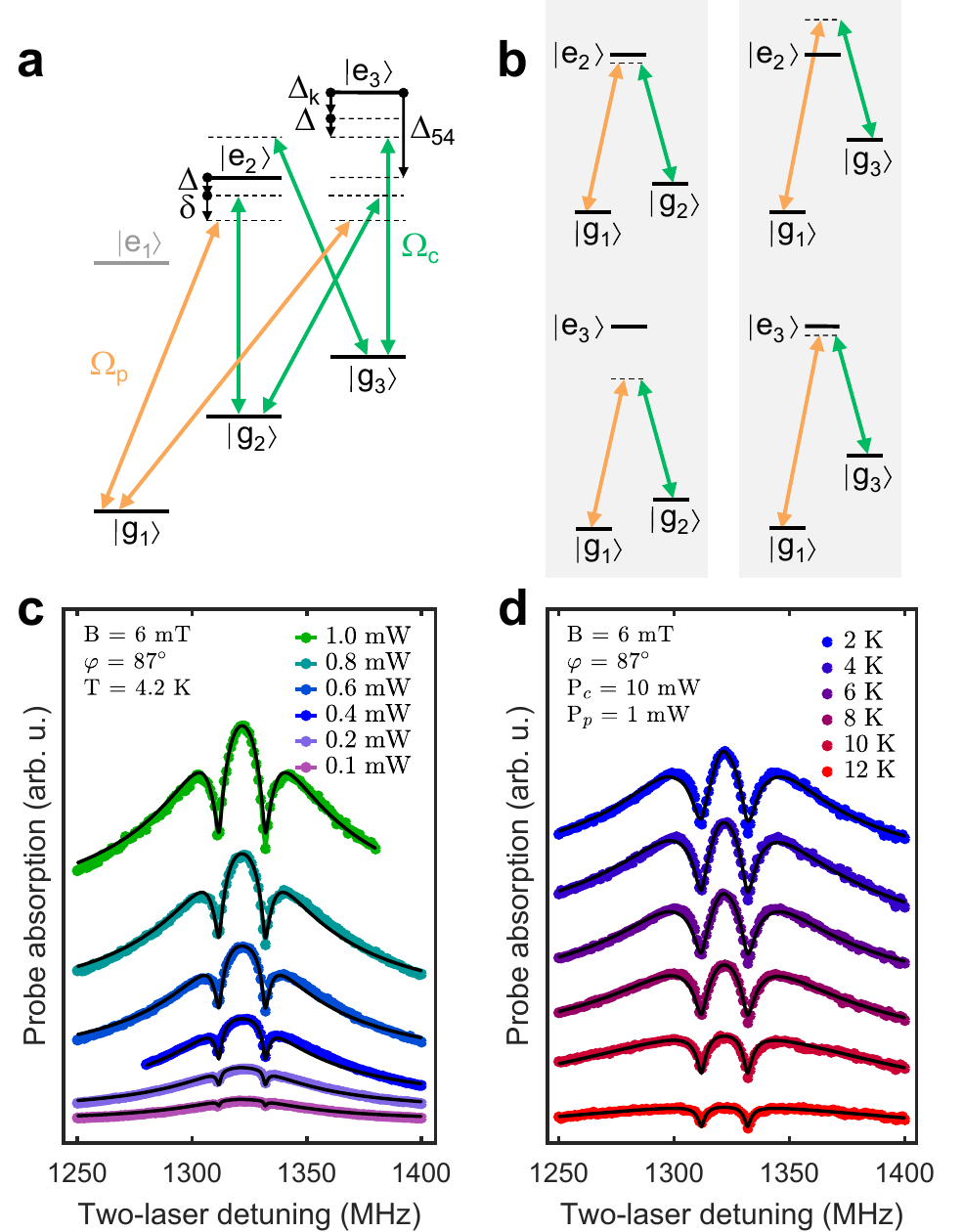}\\
	\caption{\textbf{Double EIT features in two-laser absorption spectroscopy.} \textbf{a)} At low magnetic field strength two $\Lambda$ systems are driven simultaneously. As a result, two EIT features appear within the same peak in two-laser absorption spectroscopy. One EIT feature (at $\delta$ = 0 in panel \textbf{a}) originates from interference of transitions from the $\ket{g_1}$ and $\ket{g_2}$ states (left panel in \textbf{b}). The other EIT dip appears when $\delta = \Delta_k-\Delta_{54}$ and arises from interference between transitions from the $\ket{g_1}$ and $\ket{g_3}$ states (right panel in \textbf{b}). \textbf{c)} Laser-power dependence of double EIT. Control- and probe-laser powers are varied simultaneously. The black traces in panels \textbf{b-c} are fits obtained from solving the master equation in Lindblad form for the five level system and laser couplings from panel \textbf{a} (and in the presence of 140~GHz inhomogeneous broadening). \textbf{d)} Temperature dependence of double EIT. A vertical offset has been added for clarity in \textbf{c)} and \textbf{d)}.}
	\label{fig:EIT_FigDoubleEIT}
\end{figure}

In practice, this situation arises at some external field configurations, when multiple optical transitions are possible between the spin-1 ground and optically excited states. Take for example the six-level system in Fig.~\ref{fig:EIT_FigDoubleEIT}a. Generally, in a six-level system with triplet ground and excited states a total of nine $\Lambda$ systems are possible (three sets of ground states with couplings to one of three excited states). When the ground-state spin splitting between levels $\ket{g_2}$ and $\ket{g_3}$ is comparable to the excited-state splitting between $\ket{e_2}$ and $\ket{e_3}$, and both are much smaller than the homogeneous optical linewidth of these defect centers, the control-laser couples the $\ket{g_2}$ and $\ket{g_3}$ to $\ket{e_2}$ and $\ket{e_3}$ simultaneously. If we additionally choose a configuration where $\ket{g_1}$ is sufficiently far in energy from $\ket{g_2}$, $\ket{g_3}$ (and similarly in the excited state), we ensure that the probe laser couples only to optical transition between $\ket{g_1}$ and $\ket{e_2}$, $\ket{e_3}$. In this case, only four different $\Lambda$ schemes can be driven in a single subensemble.

Figure~\ref{fig:EIT_FigDoubleEIT}b highlights these four three-level $\Lambda$ schemes, which are responsible for double EIT. The $\Lambda$ schemes that couple $\ket{g_1}$ and $\ket{g_2}$ to a shared excited state (left panel) fulfill the two-laser resonance condition necessary for observing CPT at $\delta = 0$ just as before. Additionally, the schemes that couple $\ket{g_1}$ and $\ket{g_3}$ to a shared excited state (right panel) fulfill this two-laser resonance condition at $\delta = \Delta_k - \Delta_{54}$ (see the figure for the definition of $\Delta_{54}$). In order to observe a double EIT dip the conditions need to be tuned carefully: $\Delta_k$ needs to be large enough to get off-center EIT, while at the same time the magnetic field is weak enough to have all four $\Lambda$ systems of Fig.~\ref{fig:EIT_FigDoubleEIT}b occur within the typical linewidth of the two-laser absorption peak. To achieve this, we set the angle $\varphi$ between the magnetic field and c-axis to $\ang{87}$ and tune the magnitude to 6~mT.

Figures~\ref{fig:EIT_FigDoubleEIT}c-d show the experimentally observed probe-absorption versus two-laser detuning in this configuration. Two EIT dips are visible, one on either side of the absorption peak. For Fig.~\ref{fig:EIT_FigDoubleEIT}c we varied the optical powers of both the control and probe lasers simultaneously. This allows us to extract the Rabi frequencies and dipole strengths for the various transitions accurately. We fit the traces to a model for a five-level scheme considering all transitions depicted in Fig.~\ref{fig:EIT_FigDoubleEIT}a and in the presence of 140~GHz inhomogeneous broadening (see Supplementary Information, Sec.~E). 
We get for the excited-state decay rate $\Gamma_e = (2.7 \pm 0.4)~$MHz, for the Rabi frequency $\Omega_c = 7.4~$MHz at 1 mW control beam power, and for the ensemble-averaged ground-state dephasing rate $\gamma_g^* = (0.23 \pm 0.06)~$MHz. The inverse of the last value represents the ensemble-averaged ground-state coherence time $T_2^*=(4.3 \pm 0.5)~\mu$s, which is slightly longer than what was found in microwave experiments on similar ensembles of c-axis divacancy defects \cite{koehl2011}, yet orders of magnitude shorter than the recently reported decoherence time of single divacancy defects \cite{anderson2021}. This indicates that the simultaneous driving of various optical transitions in our experiment does not significantly add to spin dephasing.

The temperature dependence of the double EIT pictures, shown in Fig.~\ref{fig:EIT_FigDoubleEIT}d for temperatures between 2 and 12~K, shows that the overall absorption peaks related to the TLAF broadens with increasing temperature, but the width of the EIT dips remains relatively stable. We fit these traces to the same model as before, but with fixed transition dipole strengths and related Rabi frequencies. From these fits we find that the ground-state dephasing rate $\gamma^*_g$ does not change significantly with temperature. In contrast, the total excited state decay and dephasing rate $\Gamma_e+\gamma_e$ does increase with increasing temperature above 6~K. We cannot determine $\Gamma_e$ and $\gamma_e$ independently from our fits. However, we know that the PL spectrum of divacancies are only moderately dependent on temperature in this range \cite{Magnusson2018}, indicating that $\Gamma_e$ also is only moderately affected by the temperature changes in our experiments. Thus, the change observed in the traces of Fig.~\ref{fig:EIT_FigDoubleEIT}d with increasing temperature is predominantly associated with the increase in the excited-state dephasing time $\gamma_e$, from $\gamma_e \ll \Gamma_e$ to $\gamma_e \approx 14 \pm 3$~MHz at 12~K. These results thus show that the ground-state dephasing time of divacancies is relatively insensitive to temperature changes in a cryogenic environment, in accordance with observation of coherent manipulation of these defects up to room-temperature. At high temperatures, the increase of the excited-state dephasing rate may limit resonant optical control and readout of these defect spins.

With our fit values for the ground-state dephasing rate $\gamma^*_g$ and the actual control-laser Rabi frequency $\Omega_c$, we can estimate the intensity requirements for complete EIT using Eq.~\ref{eq:EITlimits}. For our system with 140~GHz inhomogeneous broadening the minimum value for $\Omega_c$ should be 180~MHz, which corresponds to a control laser power of 600~mW. These optical powers could not be achieved in our experimental setup. However, higher intensities could be realized for optically thick ensembles when using single mode SiC waveguides. For example, using a divacancy-doped waveguide with a mode diameter of $5~\mu$m would yield a factor 200 larger intensity compared to our experiments. In this way, complete EIT could be achieved whilst still using tunable diode lasers in the milliwatt range. Such waveguide devices would also enhance the homogeneity of the optical fields throughout the ensemble, ensuring that all defects within the optical beam path can contribute equally to the emergence of EIT.


\section{Conclusion}

In this work we investigated EIT in a divacancy ensemble in SiC. Although its rich level structure and large inhomogeneous broadening are thought to hamper the occurrence of EIT, we show that this is not necessarily the case. In fact, we show that the inhomogeneity of the optical transition leads to interesting physical manifestation of the EIT phenomenon, and in particular unconventional EIT lineshapes that provide further information about the configuration of the defects. In multi-level systems where the eigenlevels are sensitive to external fields, the energy splittings can be engineered to allow EIT with only two lasers by selecting measurement geometries where different optical transitions are simultaneously resonant with a single laser. Slight deviations from these experimental configurations (only a few degrees misalignment for instance in the case of the data presented in Fig.~\ref{fig:EIT_FigEIT}b,c) leads to drastically different EIT dip manifestation. Combined with the fact that EIT is a purely optical phenomenon, this indicates that this technique could be used for sensing applications that are otherwise incompatible with electronic preparation of ground-state coherences. Additionally, the methods presented here for obtaining and modeling EIT control of inhomogeneous ensembles of defect spins are also valid for various other defect systems in solid-state environments, where the usual requirements of a closed three-level system with highly homogeneous transitions do not hold.

\section{Supplementary Information}
The supplementary information (available online) includes details on sample preparation (Sec.~A) and measurement geometry (Sec.~B); modelling of the magnetic-field dependence of the absorption features that allows us to identify TLAFs L\textsubscript{1}-L$_3$ (Sec.~C); a description of the Lindblad equations used to model the three-level and five-level systems described in the main text (Sec.~D); the details on the fitting routine used to include the inhomogeneous broading of the optical transitions in our model (Sec.~E); and further plots on the dependency of the inhomogeneous aborption of a three-level system and other parameters such as ground-state relaxation time (Sec.~F).

\textbf{Data availability} The data that support the findings of this study are available from the corresponding author upon reasonable request.

\textbf{Acknoweledgements} We appreciate early discussions with Danny O'Shea and Erik Janz\'en. N.T.S. acknowledges support from the Swedish Research Council (grant No. 2016--04068), the Knut and Alice Wallenberg Foundation (grant No. KAW 2018-0071). N.T.S. and C.H.v.d.W. acknoweledge support from EU H2020 project QuanTELCO (grant No. 862721). T.O. acknoweledges support from JSPS KAKENHI (21H04553 and 20H00355).

\textbf{Author contributions} O.V.Z. and C.H.v.d.W. initiated the project. O.V.Z. and X.Y. performed experiments. N.T.S. and T.O. grew and irradiated the samples. O.V.Z., T.B, A.R.O. and C.M.G. were responsible for data analysis and modelling. O.V.Z., T.B., C.M.G. and C.H.v.d.W. wrote the manuscript. All authours discussed and approved the manuscript.


\FloatBarrier
\bibliography{EIT_bib}

\end{document}